\documentclass[12pt,preprint]{aastex}

\usepackage{times}

\newcommand\oratio{\mbox{O$^{7+}$/O$^{6+}$}}

\newcommand{\ttt}{\ensuremath{TT}}
\newcommand{\trace}{{\em TRACE}}
\newcommand{\ace}{{\em ACE}}
\newcommand{\soho}{{\em SOHO}}

\newcommand{\eg}{{\em e.g.},\ }
\newcommand{\ie}{{\em i.e.},\ }

\shortauthors{Leamon \& Mc{I}ntosh}
\shorttitle{Solar Wind Forecasting from the Chromosphere}

\begin{document}

\title{Empirical Solar Wind Forecasting from the Chromosphere}

\author{R.~J. Leamon}
\affil{ADNET Systems, Inc.\ at NASA Goddard Space Flight Center, Mailcode 671.1, Greenbelt, MD 20771}
\email{leamon@grace.nascom.nasa.gov}
\and
\author{S.~W. Mc{I}ntosh}
\affil{Department of Space Studies, Southwest Research Institute, 1050 Walnut St., Suite 400, Boulder, CO 80302}
\email{mcintosh@boulder.swri.edu}

\begin{abstract}
Recently, we 
correlated the inferred structure of the solar chromospheric plasma topography with 
solar wind velocity and composition data measured at 1~AU. 
We now offer a physical justification of these relationships and present
initial results of a empirical prediction model based on them. 
While still limited by the fundamentally complex physics behind the origins of the solar wind and how its structure develops in the magnetic photosphere and expands into the heliosphere, our model provides a near continuous range of solar wind speeds and composition quantities that are simply estimated from the inferred structure of the chromosphere. 
We suggest that the derived quantities may provide input to other, more sophisticated, prediction tools or models such as those to study Coronal Mass Ejections (CME) propagation and Solar Energetic Particle (SEP) generation.
\end{abstract}

\keywords{Sun: chromosphere -- 
Sun: solar wind -- Sun: solar-terrestrial relations}

\section{Introduction}
\label{sec:intro}

%
%
Several authors have attempted to derive simple relations connecting large scale variations in observed solar quantities and plasma quantities measured {\em in situ}: 
  \cite{WangSheeley90} 
showed that solar wind speeds are anti-correlated with the super-radial expansion factors of magnetic flux tubes rooted in the solar photosphere; \cite{ArgePizzo00} developed a near-real prediction time tool based on 
  \citeauthor{WangSheeley90}'s
derivation. 
  \cite{SchrijverDeRosa03} 
developed similar predictive tools using observations from the 
{\em Solar and Heliospheric Observatory} 
\cite[\soho;][]{Fleck+Domingo+Poland1995}.
While it is widely accepted that the steady fast and slow solar wind pattern is modulated by the rotation of large open and closed magnetic structures on the solar surface, it is clear that the solar wind structure is permeated by 
much small scale structure. 

As yet, it is unknown how these small spatial structures in the solar wind originate; however, one reasonable explanation may lie in the fact that these large magnetic structures on the Sun are partitioned into even smaller magnetic structures which exist on spatial scales that are below current instrumental resolution. If the smaller structures that are embedded in the slow-fast wind are the seeds of SEP events driven by large disturbances of the wind \cite[\ie Coronal Mass Ejections or CMEs;][]{Reames99} then we need to explain, or at least quantify, the continuum of variation observed in solar wind velocities and compositions before we can reliably predict the onset of such destructive events.

We present the first results of an empirical model for prediction of solar wind conditions that is based on detailed measurements of chromospheric structure. This simple model was developed to test the measurements of \cite{McIntoshEA04} and \cite{McIntoshLeamon05} and examine why the chromospheric plasma observed by the {\em Transition Region and Coronal Explorer} \cite[TRACE;][]{Handy+others1999} appears to couple so well to that of the heliosphere measured in situ 
at 1~AU 
by the {\em Advanced Composition Explorer} \cite[ACE;][]{StoneEA98}. 
We will show that the 
empirical 
forward propagation of chromospheric structure, 
through a series of derived power-laws, allows us to recover a broad spectrum of variation in solar wind 
speed and composition 
which qualitatively reproduces that observed.

\section{The Chromospheric Footprint of the Solar Wind}
\label{sec:footprint}
\cite{McIntoshEA04} studied the ``phase travel-time'' \cite[\ttt{};][]{FinsterleEA04b} of magneto-acoustic waves observed to propagate between the 1700\AA\ and 1600\AA\ \trace\/ ultraviolet continuum pass bands and their correlation to the extrapolated chromospheric topography in an equatorial coronal hole. 
Changes in travel-time measured inside a coronal hole correlate strongly to the magnetic topography present and suggest a significant change in atmospheric conditions at the base of the chromosphere, relative to the quiet Sun. 
This single observation of possible structural changes in the chromosphere under a coronal hole sparked a broader question: Is there a significant connection between the structure of the chromosphere and that observed in the solar wind?

\cite{McIntoshLeamon05}, using very simple assumptions, showed that there was a significant connection between the inferred structure of the chromosphere and 
direct ACE measurements of solar wind conditions for a variety of 
chromospheric magnetic topologies. 
Specifically, we demonstrate that diagnostics of atmospheric ``depth'' in the chromosphere, such as the travel time of dispersive magneto-acoustic waves between the 1600\AA\ and 1700\AA\ \trace\/ UV passbands, correlate very strongly with {\em in situ\/} solar wind velocity ($V_{SW}$) and inversely with the ratio of ionic oxygen (\oratio) densities, and that a single power law relationship existed between \ttt\ and $V_{SW}$ or \oratio\ held for fast outflows from coronal holes, slow outflows from the streamer belt and intermediate outflows from quiescent sun regions. 
The best-fit power law relationships, $X = A(\ttt)^B + C$, 
are, for the wind velocity $V_{SW}$,
\begin{equation}
\begin{small}
  \begin{array}{lll}
   A=0.053 \pm 0.007; &
   B = 4.56 \pm 0.33; &
   C = 333 \pm 12
  \end{array}
  \label{eqn:vsw}
  \end{small}
\end{equation}
%
%
%
%
and for the composition ratio \oratio{},
\begin{equation}
\begin{small}
  \begin{array}{lll}
   A = 29.9 (\pm 2.70) \times 10^{3}; \\
   B = -7.21 \pm 0.23; &
   C = 0.011 \pm 0.003.
  \end{array}
  \label{eqn:o76}
  \end{small}
\end{equation}

Even before it enters the corona, the solar gas is fractionated; transport through the chromosphere into the corona causes an overabundance of elements with low first ionisation potential such as magnesium 
(the so-called FIP effect). \cite{GeissEA95} studied the variations in Mg/O and \oratio\ during Ulysses' 1992-93 passages through the southern High Speed Stream (\ie Coronal Hole), and found that they were highly correlated. Since the Mg/O ratio is controlled by the FIP effect and the \oratio ratio reflects the coronal temperature, the correlation between the two points to a connection between coronal conditions and the structure of the chromosphere. We find that Mg/O is also highly correlated with \ttt{}.
\cite{GeissEA95} further showed that, as the Sun 
(and the coronal hole) 
rotated under Ulysses, the transition between the coronal hole and quiet sun was almost 
as steep as 
a step function. Thus the chromosphere and corona have a common, relatively sharp, boundary, separating the low-FIP from the high-FIP region in the chromosphere and the low-$T$ from the high-$T$ region in the corona. The existence of such a common boundary points to a causal relationship of the kind for which conditions in the chromosphere determine the supply of energy into the corona.

That coronal holes give rise to fast streams of wind depleted in hot oxygen is not a new result;
the key is that there is one continuous function describing a range of solar wind speeds derived from the structure of the chromosphere, just as there is a continuous energy input spectrum responsible for the plasma topology. 

\subsection[How the footprint comes to be: Energy input into the solar atmosphere]
{How the footprint comes to be: \\Energy input into the solar atmosphere}
\label{sec:SUMER}


\cite{McIntosh2006b} offer an explanation of atmospheric heating and initial solar wind acceleration (inside and outside of coronal holes respectively) that is observationally consistent with a single physical mechanism being responsible for both processes.
The correlations between different spectroscopic features observed 
are consistent with the energy input to the plasma via the action of relentless magnetoconvection-driven magnetic reconnection; the energy input to the plasma is modulated by the unsigned field strength on the supergranular length scale while the closure of the global magnetic topology determines if the injected mass and energy heats closed loops or is accelerated along open field lines.

In a coronal hole, the small (spatial) scale flux elements anchored in the super-granular boundary  are effectively open to interplanetary space because they have the same magnetic polarity as the bulk of the coronal hole. As a small, recently emerged magnetic dipole is advected to the boundary, the leading polarity of the flux begins to reconnect with the anchored element, creating a new magnetic topology in the super-granule interior. A portion of the energy released by the reconnection quickly begins to evaporate cool chromospheric material into the topology whereas the bulk of the remaining energy is released in the form of kinetic energy. 

Conversely, in the quiet Sun
the magnetic flux anchored in the super-granule boundary  closes somewhere in the vicinity of the cell with a similarly anchored, opposite polarity piece of magnetic flux on a nearby super-granule, such that there is very little probability that any arcade created in the super-granule interior can open into interplanetary space. Thus, as the reconnection progresses, the bulk of the released energy must result in the evaporative mass-loading of the created arcade and thermal heating of the plasma contained therein, since little of it can be rapidly converted into plasma outflow. 


Section~5.1 of \cite{McIntosh2006b} discusses the likelihood that the energy input mechanism and its connection to the global magnetic topology are tied to Geiss' observation of the FIP effect. We prescribe that the fast solar wind generally originates from open magnetic regions \cite[with sizable supergranular magnetic imbalance,][]{McIntosh2006a} and that the energy is delivered in a mostly non-thermal (kinetic) form with mass that originates in the well mixed chromosphere. Conversely, the slow solar wind originates from a largely closed magnetic topology in the quiet Sun, which will ensure a complex, mostly thermal, energy delivery to the plasma. It is possible that enhanced fractionation can take place in the plasma as it is transported through the magnetic topology and eventually evaporated out of the corona and into the heliosphere in a fashion that depends on the exact magnetic topology of the observed plasma \cite[e.g.,][and references therein]{SchwadronMcComas03}.

Many other markers of the coupling between the initial release and distribution of energy and the (supergranular) magnetic flux balance of the solar plasma may have already been observed. For example, systematic He~{\sc i} 10830\AA\ line asymmetries (observed in the chromosphere at the solar South pole near solar minimum) 
have been attributed to a strong outflow (\mbox{$\sim$8 km s}$^{-1}$) of chromospheric material \cite[]{Dupree1996} and occur at a time when there is a relatively strong unbalanced field at the pole. The magnetic flux that diffuses to the polar regions over the course of the solar cycle creates a long-lived ``polar crown'' coronal hole at solar minimum. This excessive imbalance in the polar magnetic field, we suspect, acts as a reservoir of kinetic energy for the very high speed winds observed at high heliospheric latitudes at the same phase of the solar cycle \cite[]{McComas1998} while the evolving mixture of open and closed regions and the resulting energy partitioning of the plasma explain the complex heliospheric structure observed at other phases of the solar cycle \cite[e.g.,][]{McComas2000, Smith2003}.

The above discussion offers support to, and an energy input mechanism for, two recent solar wind models that relate solar wind outflow speed at 1~AU (inversely) to coronal temperatures, and that energy input is shared between thermal energy and wind energy in one consistent manner \cite[]{Fisk03,SchwadronMcComas03}. Further, it justifies our hypothesis that we can use observations of the chromosphere as proxies for plasma conditions in the corona and solar wind, even if (as) the perceived
direct driver of the solar wind is expansion of the million-degree 
corona high above the chromosphere.
%
The eventual energy balance {\em must\/} be reflected in the travel-time diagnostic 
\trace\/ observations of the solar chromosphere in corona holes and quiet Sun through the visible changes in the plasma topography.



\section{Forecasting the Solar Wind from the Chromosphere: Method}
\label{sec:method}

If we can correlate the wind speed and composition to the conditions in the solar chromosphere, can we use the correlations to predict conditions in the solar wind? In principle, the answer is yes, but the nature of the \trace\/ UV observations does not lend itself to a predictive tool: each of the 13 solar observations (data points) in  \cite{McIntoshLeamon05} is the result of a 1--3 hour time \trace\/ sequence of interleaved 1600\AA\ and 1700\AA\ images, which require careful co-alignment to sub-pixel accuracy and removal of the effects of solar rotation. Fortunately, there is a striking relationship  \cite[]{FinsterleEA04b, McIntoshEA04}  between the observed travel times and the plasma $\beta$ (ratio of gas pressure to magnetic pressure) in the \trace\/ field of view at the (presumed) mean formation height of the \trace\/ 1600\AA\ passband  \cite[some 450~km above the photosphere;][]{FossumCarlsson05}. 
Extrapolating a co-aligned line-of-sight magnetogram from the Michelson Doppler Imager \cite[MDI;][]{ScherrerEA95}  on \soho{} to 450~km, computing the magnetic pressure ($= |B|^2 / 8\pi$) and by imposing a simple model gas pressure  \cite[][model VAL3C]{VernazzaEA81},  we can compute $\beta$ in the chromospheric region of the \trace\/ passbands.\footnote{The \ion{Ni}{1} line at 6768\AA\ used by MDI is itself formed some 200~km above the photosphere, so we are, in fact, only extrapolating the magnetic field 250~km.}

\begin{figure}
\epsscale{0.85}
\plotone{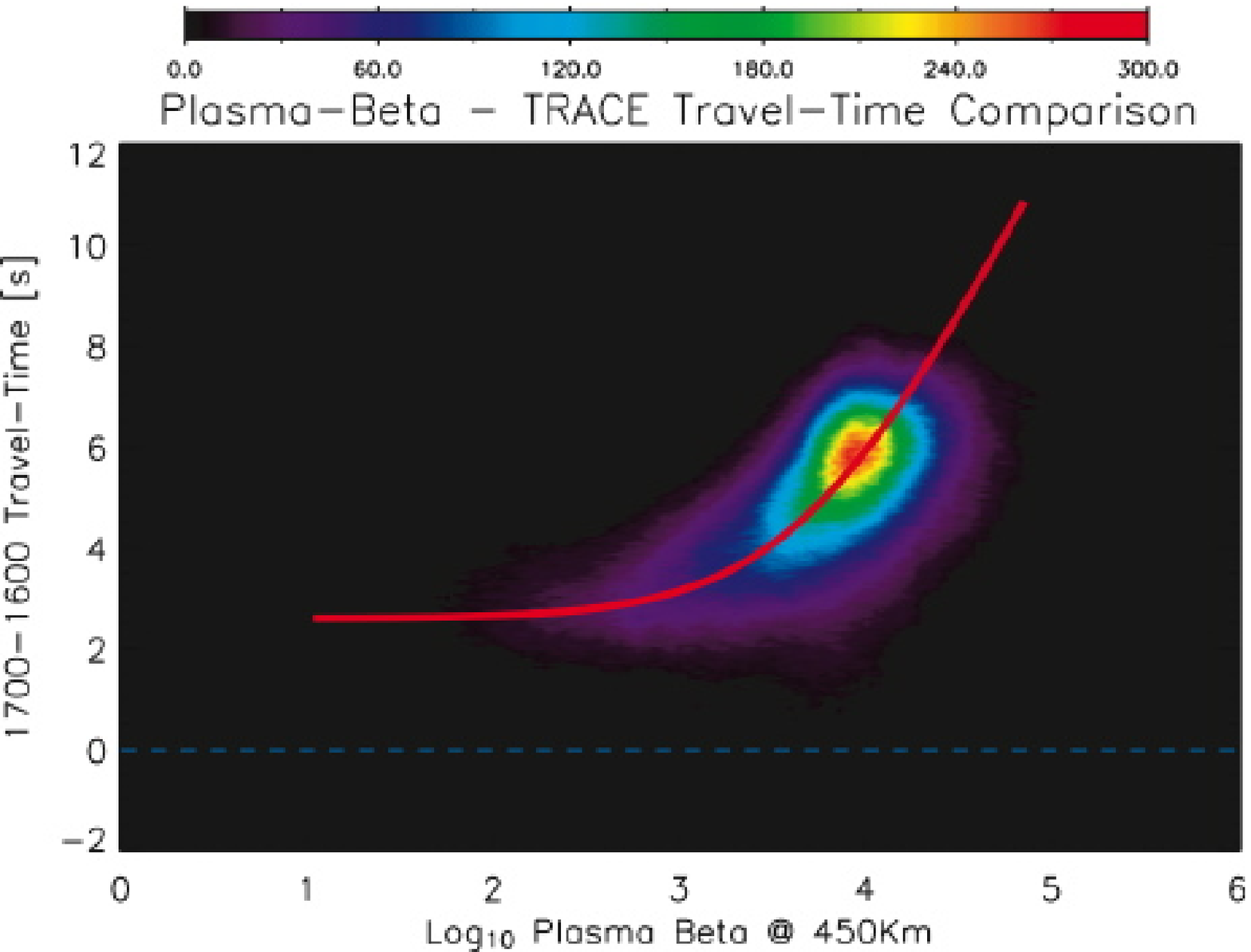}
\caption{2-D histogram showing the relationship between plasma $\beta$ at 450~km and \trace\/ UV travel time, for all pixels in all of the 13 events studied by \protect\cite{McIntoshLeamon05}. The bins are 0.05 wide in both $TT$ and $\log_{10} \beta$, and the color scale is in pixels per bin.
The best fit curve is \mbox{$ TT = -23.3 + 7.0 \log_{10}(\beta + 5 \cdot 10^3)$}.
} 
\label{fig:scatter}
\end{figure}

\begin{figure}
\plotone{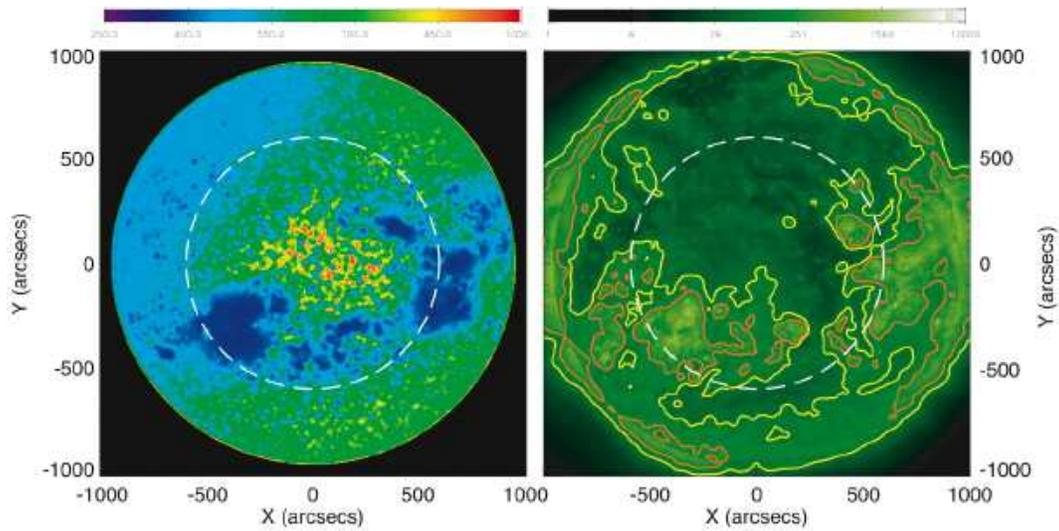}  
\caption{(left) Predicted full-disk solar wind speed map, as inferred from the 2003 August~6, 03:15~UT \soho\ MDI magnetogram using the $\beta$--$TT$ relationship shown in Fig.~\protect\ref{fig:scatter} and the $TT$--$V_{SW}$ relationship of Eq.~(\protect\ref{eqn:vsw}).
(right) \soho\ EIT context image from  2003 August~6,  03:12~UT. The yellow and orange contours show the 100 and 200 DN intensity levels in the image, respectively \protect\cite[cf.][]{McIntoshEA04}. In both panels, the white dashed circle has radius 600\arcsec, where the accuracy of the line-of-sight approximation for magnetograms degrades significantly.
There are visible artifacts beyond this perimeter, especially towards the northeast and southwest.
}
\label{fig:ett}
\end{figure}

Fig.~\ref{fig:scatter} is a two-dimensional histogram that shows the relationship between the observed oscillation travel time and the plasma $\beta$ at 450~km above the photosphere for the suite of 13  \trace\/ observations presented in \cite{McIntoshLeamon05}. 
%
Using the $\beta$--\ttt{} 
relationship, we can generate a synthetic full-disk \ttt{} map from the extrapolated MDI magnetogram $\beta$ and then use the power law of  Eqn.~(\ref{eqn:vsw}) to generate a map of wind speed or other correlated in situ quantities. 
{Fig.~\ref{fig:ett}} shows examples the derived wind speed map, along with a context EUV coronal image from the \soho{} Extreme-Ultraviolet Imaging Telescope \cite[EIT;][]{Boudine1995}. The latter clearly shows the active regions and coronal holes present on the disk. 

To demonstrate our method, we compare our predictions to observations over the course of approximately one whole solar rotation, from July~11 to August~10, 2003. This interval was chosen to overlap with the real TRACE 
UV observations made on July~14, 20, and~27 \cite[see][]{McIntoshLeamon05}.

Although we compute the in situ predictions for the whole Sun, we average them over an Earth-directed ``source window,'' 
created by tracing
33 field lines back from the $2.5 R_\sun$ source surface of a Potential Field--Source Surface (PFSS) model to the photosphere.
{Fig.~\ref{fig:pfss}} shows PFSS field-line tracings for 4 times 
(separated by approximately one week)
spread through our sample interval. 
The angular distance between the point the sub-terrestrial field line maps back to and the sub-terrestrial point ranges between zero (\ie maps straight back) and as much as 35\degr\ away from disk center; the average separation is 14\degr{}.
The other 32 field lines are arranged on the perimeter of an ellipse with semi-major axes $5\degr \times 2.5\degr$. These 32 field lines usually map back to a small closed region on the photosphere. The 4th panel of Fig.~\ref{fig:pfss} shows a time when this is not the case; the ellipse maps to a crescent-shaped swath covering most of the northern solar hemisphere.

To calculate the time it takes a plasma parcel to propagate out to 1~AU, we use the method of \cite{Cranmer04},
invoking Lambert's $W$ function, to solve the original solar wind model of \cite{Parker58a}.
With 
the predicted wind speed at 
1~AU as an initial condition, 
we integrate the time-of-flight backwards to the sun.  
Adding the time-of-flight to the MDI observation dates gives a time-series in situ prediction, which we compare to \ace\ data.
Interestingly, the time difference between ballistic and accelerating flows arriving at 1~AU scales as $V_{SW}^{-3/2}$.

\section{Results and Discussion}
\label{sec:results}

\begin{figure}
\epsscale{0.85}
\plotone{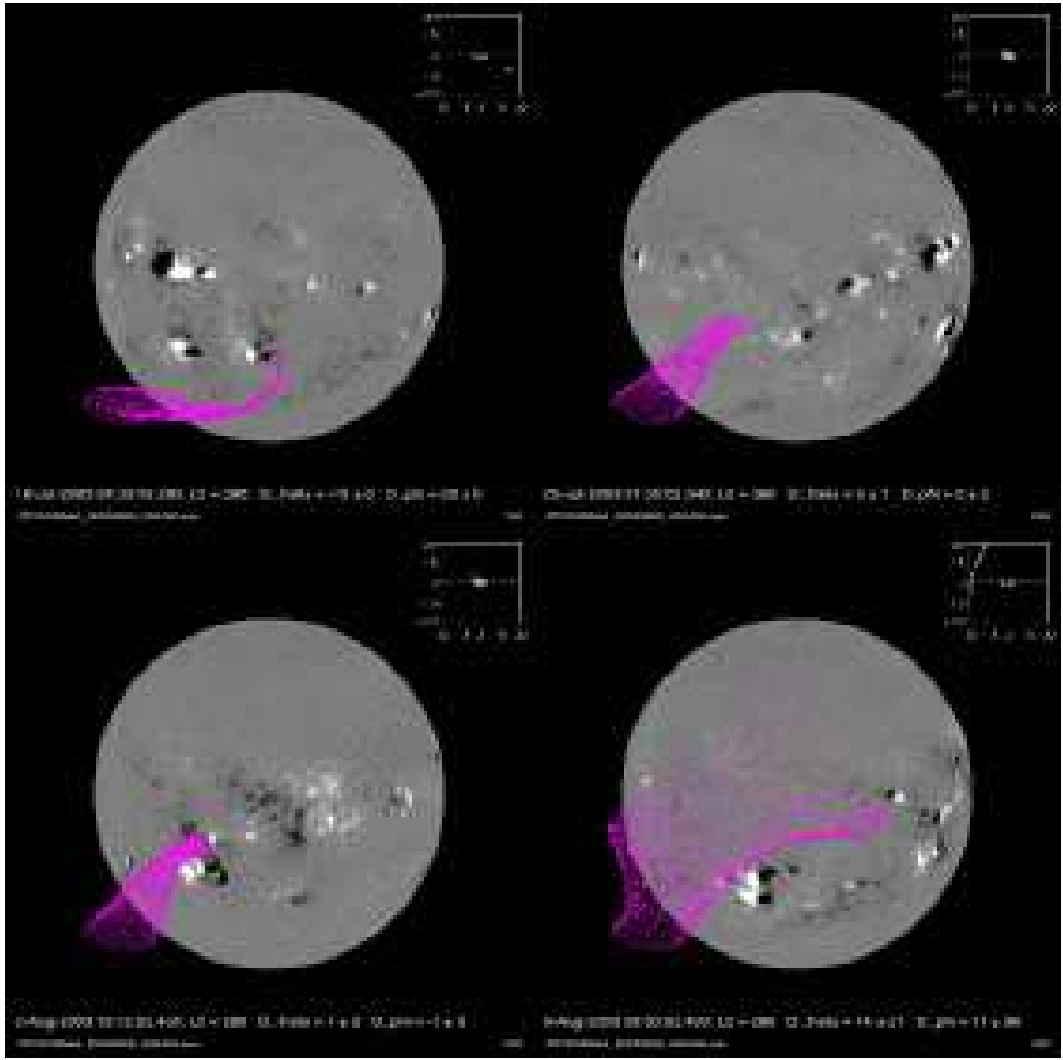}
\caption{Tracing the sub-terrestrial field lines back to the photosphere. For each MDI magnetogram, 33 field lines are traced back to the sun; one at the sub-solar point, and 32 around an ellipse $5\degr \times 2.5\degr$. Sometimes, the whole ellipse maps to a small region of the photosphere, other times it maps to regions all over the photosphere. Sometimes the sub-terrestrial point on the source surface is connected to the sub-terrestrial point on the photosphere, other times the footpoint is as much as $35\degr$ away from disk center. [{\em See the electronic edition of the Journal for an mpeg animation of this figure.}] 
} 
\label{fig:pfss}
\end{figure}

\begin{figure}
\epsscale{0.65}
\plotone{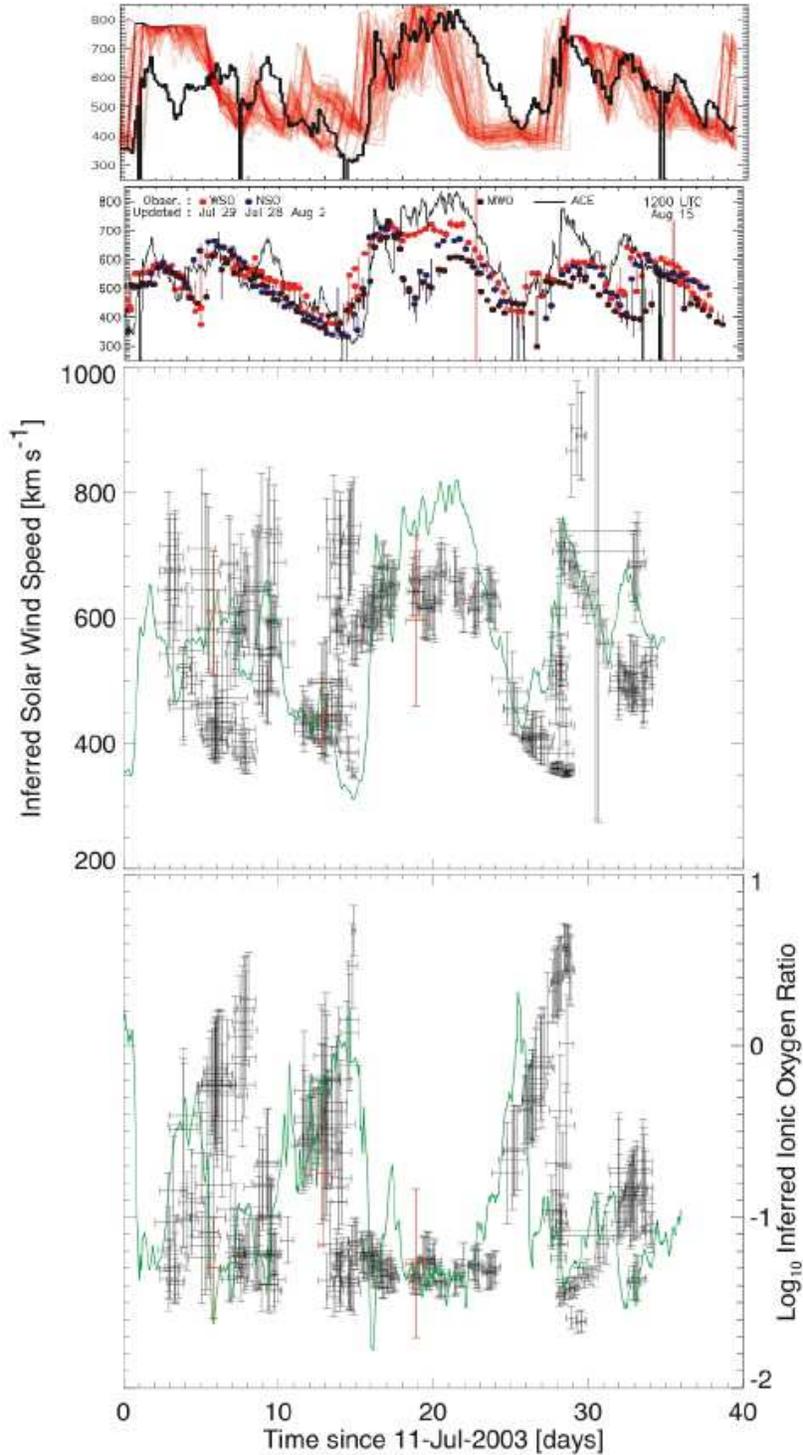}
\caption{(lower two panels) Predicted solar wind speed and oxygen composition \oratio\ (bottom) extrapolated from inferred $TT$ and propagated out to 1~AU\@. Quantities as observed by \ace\/ are in green, and the three orange points are real \trace\/ \ttt\ observations. 
(upper two panels) Predicted wind speed from 
DeRosa \&\ Schrijver
(above; {\tt http://www.lmsal/forecast/})
and
Wang, Sheeley \&\ Arge
(below; {\tt http://www.sec.noaa.gov/ws/}). 
} 
\label{fig:results}
\end{figure}

The lowest two panels of
{Fig.~\ref{fig:results}} show the results of our propagation model. For comparison purposes, the TRACE observations mentioned above are marked on Fig.~\ref{fig:results} in orange. 
There is a good correlation,
albeit with the predictions lagging the observations. For the whole interval, the peak cross-correlation coefficient of 0.47 occurs at a lag of +19.6h for \oratio, and 0.32 at +24h for $V_{SW}$. The Spearman rank-order correlation coefficients are somewhat higher: 0.63 and 0.46, respectively.
We attribute the long lags to the unusually slow wind speeds resulting from periods when the wind emanates from active regions. Not unrelated is the variation in error bar size---smallest when the whole ellipse maps back to one point in an active region (\ie slowest wind), and large (\eg panel~4 of Fig.~\ref{fig:pfss}).  
This is not surprising, considering that the \trace\/ data used to derive Eq.~(\ref{eqn:vsw}) are averaged over a $200'' \times 200''$ region and contain a range of different \ttt\ values, whereas the smallest PFSS footprints are approximately
$80''$ square. It is clear that a sub-field less than a quarter the size of that of \cite{McIntoshLeamon05} will derive more extreme values of \ttt{} and adjust the power-law relationships accordingly.

The top panel of Fig.~\ref{fig:results} shows the predictions for the same time interval of the LMSAL DeRosa-Schrijver model 
and the second panel those of the NOAA SEC Wang-Sheeley-Arge model. 
The 
LMSAL forecast is in fact 25 different predictions (red traces), from 25 slightly different initial conditions (varying base wind speed, stream interaction strength, etc.).
Similarly, the 
WSA model is uses data from three different sources as inputs. 
Much has been made of the intercalibration 
between the Mount Wilson (MWO), Wilcox (WSO) and Kitt Peak (NSO) data used, but it is not germane to discuss them here. 
It is reassuring to see that our model, based on chromospheric \ttt, gives essentially similar results to the other predictions, 
where the wind speed is related to the super-radial expansion factor of at each point on the Source Surface. A visual inspection suggests that our model most closely tracks the WSO-derived predictions. For such a simple model, the correspondence with both observations and more complex models is both striking and encouraging.
We cannot quote correlation coefficients for the other models for the interval studied, but the WSA method paper \cite[]{ArgePizzo00} cites that for the three-year period centered about the May~1996 solar minimum, the correlation coefficient for $V_{SW}$ is 0.4 with an average fractional deviation of  0.15. We must be happy with 0.47 and 0.32, respectively, for one month's worth of predictions given the greater variability of the solar wind in the declining phase of a solar cycle \cite[]{McComas2000}.

We have previously reported on the correlations that exist between {\em in situ\/} solar wind quantities measured by \ace\/ and diagnostics of the chromospheric plasma topography from \trace\/ observations. 
The two main results of the present work are: (1) the chromospheric plasma $\beta$ map derived from \soho\/ MDI data can be used as a useful full-disk proxy for \trace\/ travel-time measurements of chromospheric structure; and (2) the plasma $\beta$ maps can be used generate accurate forward-modelling predictions of the {\em in situ\/} solar wind state, with correct magnitudes and prediction of gross features and these predictions can be updated at the cadence of the full-disk magnetograms.
We emphasize again that this is a very simple, empirical model connecting observed changes in chromospheric structure to their impact on the structure of the solar wind. 
Nevertheless, as a first generation model (over a limited time interval), we 
obtain
a reasonable amount of detail on the structure of the nascent solar wind. 
One goal of future work is clearly the development of a real-time model.
On the other hand, perhaps we are not as truly empirical as we suggest:
We deviate from existing wind models by predicting composition.
While \oratio\ might not be a priority for space weather, that we can predict it with the same accuracy as $V_{SW}$ affirms the core concepts and physics behind the chromosphere-corona coupling outlined in 
section~\ref{sec:SUMER}.

Finally, several limitations to our model, including the using the quiet-Sun VAL3C atmospheric model everywhere and using a potential extrapolation to generate maps of $\beta$ and \ttt{},
can be simply addressed by using full disk time-series observations of the chromosphere to compute the travel-time maps without having to perform the present limited field-of-view to full disk transform. Full disk observations of this type will be a standard ``data product'' of the Solar Dynamics Observatory.

%
%

\acknowledgments

SWM's contribution is based upon work supported by 
NASA
under Grants NNG05GM75G, issued under the Sun-Earth Connection Guest Investigator Program and NNG06GC89G, issued under the Solar \&\ Heliospheric Physics Program and by the 
NSF
through its Solar Terrestrial Physics Program (ATM-0541567).

%
%


\clearpage

\clearpage

\clearpage

\clearpage

\end{document}